# Enhanced figure of merit in nanostructured (Bi,Sb)$_2$Te$_3$ with optimized composition, prepared by a straightforward arc-melting procedure

F. Serrano-Sánchez,[1] M. Gharsallah,[2] N.M. Nemes,[3*] N. Biskup[3], M. Varela[3,4], J.L. Martínez,[5] M.T. Fernández-Díaz,[6] J.A. Alonso[1]

[1]Instituto de Ciencia de Materiales de Madrid, C.S.I.C., Cantoblanco, E-28049 Madrid, Spain
[2]Sfax University, National School of Engineers, B. P. W 3038, Sfax, Tunisia
[3]Departamento de Física de Materiales, Universidad Complutense de Madrid, E-28040 Madrid, Spain
[4]Instituto Pluridisciplinar & Instituto de Magnetismo Aplicado, Universidad Complutense de Madrid, E-28040 Madrid, Spain
[5]ESS Bilbao. Pol. Ugaldeguren III, Pol.A-7B. Zamudio, E-48170 Spain
[6]Institut Laue Langevin, BP 156X, F-38042, Grenoble, France

Sb-doped Bi$_2$Te$_3$ is known since the 1950s as the best thermoelectric material for near-room temperature operation. Improvements in material performance are expected from nanostructuring procedures. We present a straightforward and fast method to synthesize already nanostructured pellets that show an enhanced ZT due to a remarkably low thermal conductivity and unusually high Seebeck coefficient for a nominal composition optimized for arc-melting: Bi$_{0.35}$Sb$_{1.65}$Te$_3$. We provide a detailed structural analysis of the Bi$_{2-x}$Sb$_x$Te$_3$ series (0≤x≤2) based on neutron powder diffraction as a function of composition and temperature that reveals the important role played by atomic vibrations. Arc-melting produces layered platelets with less than 50 nm-thick sheets. The low thermal conductivity is attributed to the phonon scattering at the grain boundaries of the nanosheets. This is a fast and cost-effective production method of highly efficient thermoelectric materials.

*) Corresponding author. Electronic mail: nmnemes@fis.ucm.es

Thermoelectric materials are able to convert temperature differences into electrical power, mainly through the scavenging of waste heat, integrated into thermoelectric generators.[1,2] Among thermoelectric materials, for near-room-temperature applications, $Bi_2Te_3$ alloys have proved to exhibit the best thermoelectric efficiency for n- and p-type thermoelectric systems.[3–8] The customary way to compare thermoelectric materials is in terms of the figure of merit, ZT, defined as ZT= $S^2\sigma T/\kappa$, (S: Seebeck coefficient, σ: electrical conductivity, κ: thermal conductivity, and T: absolute temperature). Alloying $Sb_2Te_3$ and $Bi_2Se_3$ allows for the fine tuning of the carrier concentration, along with a reduction in lattice thermal conductivity. Owing to several studies on single and polycrystalline materials, the link between electronic transport properties and dopant concentration is now understood.[9]

Highly efficient thermoelectric energy conversion could be forthcoming based on nanostructured thermoelectric materials.[10,11] Bulk samples containing nanoscale constituents exhibit enhanced properties that are relevant for optimizing the thermoelectric figure of merit. Among emerging nanostructured materials, thermoelectric nanowires received substantial attention from several groups.[12–15] Multilayered thin films were found to be another avenue to increase the figure of merit.[16]

Usually, nanostructuration of thermoelectric materials is managed in three steps: synthesis, formation of nanoparticles and their assembly into bulk solids. Several methods are employed in the elaboration of nanostructured bulk materials; the most frequently used are spark-plasma-sintering (SPS), hot pressing, ball milling and wet chemical reactions.[17] Despite the advantages shown by each synthesis method, there is the shared drawback of long reaction and sample preparation times. For example, the SPS process presents the benefit of retaining low dimensional grains, thus decreasing the thermal conductivity; on the other hand, it requires long annealing times and is also expected to result in more pronounced equiaxed morphology of the powder particles with a decrease in their size.[18] Ligand-assisted chemical methods are useful to obtain good particle size, shape and crystallinity. Nevertheless, the insulating organic capping ligands must be completely removed from the nanocrystals before bulk pellets can be formed. The ZT values of most chemically prepared materials are low, affected by inappropriate carrier concentrations and lousy intergranular connectivity.[19,20]

Several efforts have been made to improve the thermoelectric performance of well-known Bi-Sb-Te based alloys, the ones used in most commercial devices. High ZT values have been reported in superlattice structures $Bi_2Te_3/Sb_2Te_3$, but their applicability in energy conversion is limited[11]. The main advantage of the superlattice clusters is the improved thermal conductivity. Pettes *et al.* assessed the thermoelectric performance for $BiSbTe_3$ stoichiometry in single $(Bi_{1-x}Sb_x)_2Te_3$ nanoplates, which showed extremely low values of thermal conductivity ranging from 0.4 to 1.0 $Wm^{-1}K^{-1}$ and a figure of merit of 0.30.[21] Many efforts have been made in order to improve thermal conductivity and thermoelectric performance in bulk nanocomposites through different elaboration methods. Nanocrystalline Bi-Sb-Te alloys synthesized by ball milling and hot-pressing have shown ZT as high as 1.4.[22] In order to reduce bulk thermal conductivity, bismuth-antimony-telluride alloys were synthesized from their oxide reagents in a high temperature melting and reduction process, which allows control of microstructure morphology, reaching values of ZT=0.7 for $Bi_{0.4}Sb_{1.6}Te_3$.[23] Zhang *et al.* achieved a zT=0.51 in optimized $Bi_{0.5}Sb_{1.5}Te_3$ by a controlled synthesis of nanoplatelets and its sintering through SPS to form bulk nanocomposites, as a scalable bottom-up process.[24] Luo *et al.* reported on one of the highest figure of merit of 1.71 in $Bi_{0.5}Sb_{1.5}Te_3$ obtained by the optimization of traditional melting-solidification method under a variable intensity magnetic field, and described a complete study of this method.[25] An improvement in average ZT reaching 1.18 in the temperature range of 300-480 K was achieved by Hu *et al.* displaying a peak ZT=1.3 measured in $Bi_{0.3}Sb_{1.7}Te_3$ due to the resulting morphology of the samples prepared by melting and hot-deformation method.[26] The best performance is reported in Tellurium-excess melt-spun (Te-MS) samples that present the highest ZT of 1.86 at 320 K for Bi-Sb-Te alloys reported until now.[27]

In this report we describe the synthesis of $Bi_{2-x}Sb_xTe$ compounds (0≤x≤2) by arc melting. The really short reaction times of this technique yield strongly oriented polycrystalline pellets as observed by SEM and TEM, with enhanced thermoelectric properties as a result of the nanostructuration.[28,29] We describe thermal transport properties (Seebeck-coefficient, electrical and thermal conductivity, Hall-effect) and detailed atomic crystal structure determined from neutron powder diffraction (NPD); with important insights about anharmonic bonding, relevant for lower thermal conductivity, while maintaining high electrical conductivity.

## Results and Discussion

*Crystal structure*

For all Sb-alloyed $Bi_2Te_3$ compounds, we identified a $Bi_2Te_3$-type structure, defined in the space group *R-3m*, using x-ray diffraction (XRD). The as-grown sample is strongly textured, as shown by the orientation enhanced (001) reflections in the XRD diagrams (Fig. 1a). Nevertheless, we could not improve the profile fit even with a preferred-orientation function, nor explain some reflections, as indicated. Therefore, we turned to NPD to fully characterize the structural details of the $Bi_{2-x}Sb_xTe_3$ system. NPD offers many advantages for bulk diffraction studies: the large amount of powder sample and the rotating sample holder minimize the preferred orientation effects; much broader reciprocal space is available than with XRD, and the lack of form factor helps us determine the anisotropic displacement factors precisely. We refined the crystal structures in the $Bi_2Te_3$-type model[30] in the rhombohedral *R-3m* space group (no. 12), hexagonal setting, with Bi and Sb atoms distributed at random over *6c* (0 0 z) positions and the two types of tellurium, Te1 at *3a* (0 0 0) positions and Te2 at *6c*. The unit-cell parameters and main interatomic distances for the five compositions are given in Table 1; the full structural parameters for nominal x= 0.0, 1.0, 1.5, 1.65 and 2.0 are included in the Supplementary Tables S1-S5, respectively. Tables S6-S8 contain the structural parameters for x=1.65 studied at elevated temperatures. The data for $Bi_2Te_3$ have been published elsewhere but are repeated here for reference.[29]

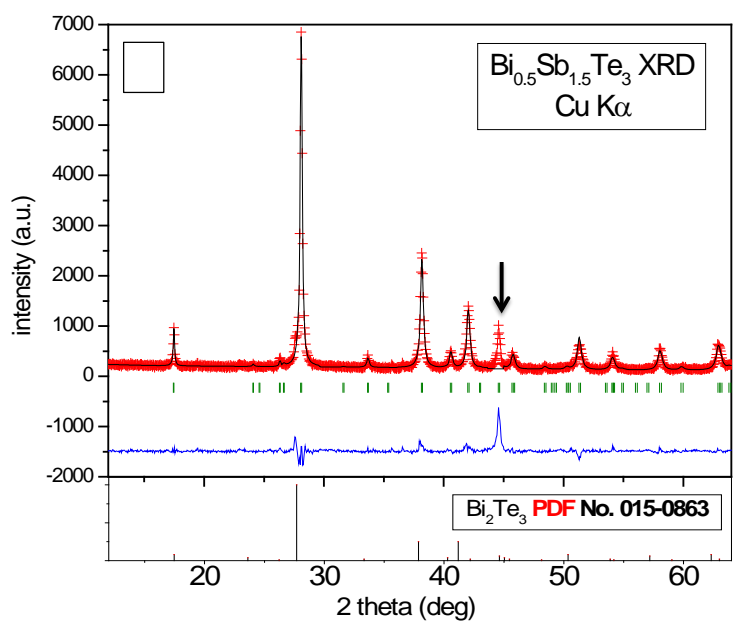

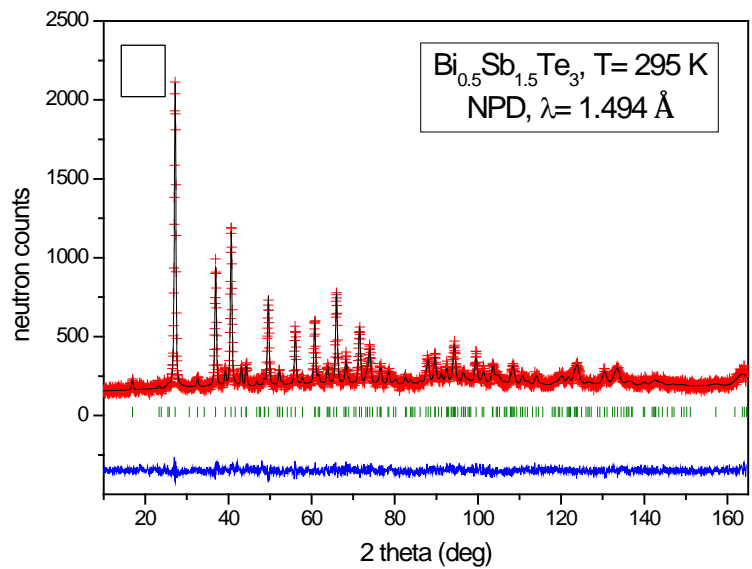

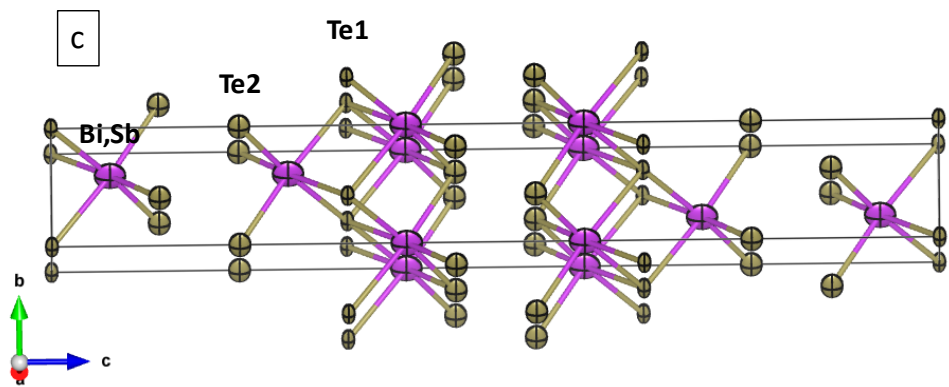

*Fig. 1. Structure and morphology of $Bi_{2-x}Sb_xTe_3$ (data shown for x=1.5) (a) XRD and (b) NPD patterns at room temperature for the same sample. Rietveld-refinement in the space group R-3m of the XRD is hindered by strong preferred orientation that enhances the [00l] reflections (see for example the [0015] reflection). The NPD pattern is well refined with observed (crosses), calculated (full line) and difference (at the bottom) profiles. (c) These compounds have layered crystal structure. The anisotropic displacement ellipsoids (drawn with 95% probability) are flattened along the [110] direction, as determined from NPD Rietveld refinement.*

In $Bi_{2-x}Sb_xTe_3$ (x= 1.0, 1.5) the occupancy factors of Bi *vs* Sb (we fixed the Te at $6c$ positions to unity) deviate somewhat from the initial stoichiometry. The actual compositions of the samples were assessed by the refinement of these occupancy factors from NPD data, which is a powerful microscopic analysis tool, based on the scattering length contrast between Bi and Sb. This analysis indeed showed a partial loss of Sb during the synthesis. The actual composition of the arc-melted samples with nominal composition x=1.0, 1.5, 1.65 was refined as x=0.66(5), 1.12(5), 1.46(3) from NPD data, respectively. The goodness of this procedure has been checked by chemical analysis by ICP-AES, yielding x= 1.49(1) for the nominal x= 1.65 composition, and also by local energy-dispersive x-ray spectroscopy (EDX) analysis of a few selected grains in the transmission electron microscope (TEM) study, finding compositions consistent with $Bi_{0.5}Sb_{1.5}Te_3$. The agreement between observed and calculated NPD profiles is excellent, (Fig. 1b for x= 1.5), and was further improved for all reflections by using a minor preferred orientation correction. The Rietveld plots for $BiSbTe_3$ and $Sb_2Te_3$ are also included in the Supplementary Figures S1 and S2, respectively.

The layered crystal structure, refined from NPD data, is shown for x= 1.5 in Fig. 1c. The layers consist of five-fold stacked, covalently bonded Te2-(Bi,Sb)-Te1-(Bi,Sb)-Te2 atoms, with van der Waals-type interatomic forces between adjacent layers (Te2-Te2 interactions). All the atoms show flat anisotropic thermal ellipsoids perpendicular to the bonding direction. The coordination of Te1 and Te2 atoms are radically different, with Te1 6-fold coordinated to (Bi, Sb) atoms 3.204(4) Å away, and terminal Te2 covalently bonded to 3 atoms at 3.000(4) Å, with their non-bonding electrons reaching into the space between layers. The Bi (or Sb) atoms are surrounded by 6 Te (3 Te1 and 3 Te2) atoms in octahedral coordination. NPD can reveal the thermal motion of the atoms, indicated by displacement ellipsoids and these are considerably larger for Te2 than for Te1, indicating a higher lability or mobility. On the other hand, compared with the structure of the parent $Bi_2Te_3$ compound, we observe a significant contraction of the unit-cell volume with increasing Sb content (e.g. $a$= 4.3008(2), $c$= 30.5006(17) Å and V= 488.58(4) Å$^3$ for the $Bi_{0.5}Sb_{1.5}Te_3$ compound, in contrast with $a$= 4.3849 Å, $c$= 30.4971 Å and V= 507.82 Å$^3$ for $Bi_2Te_3$,[29] which is a consequence of the substantially smaller ionic size of octahedrally coordinated $^{VI}Sb^{3+}$ (0.76 Å)[31] $vs$ $^{VI}Bi^{3+}$ (1.03 Å). While the volume contraction concerns the $ab$ plane, there is a slight expansion along $c$. Clearly, the replacement of $Bi^{3+}$ by smaller $Sb^{3+}$ affects mainly the cationic packing within the layers, whereas the spacing between the five-fold layers is determined by Te2-Te2 long distances (3.70 Å), which are not influenced by the substitution.

*Atomic thermal motion*

The anisotropic displacement parameters (ADP) can reveal details about the atomic vibration, relevant for decreased thermal conductivity, as they indicate strongly anisotropic motions for all three constituent sites (Bi/Sb, Te1, Te2) as shown by the corresponding thermal ellipsoids in Fig. 2a and in Supplementary Tables S1-S8. The ADP of the six-fold coordinated intralayer Te1 is approximately four times smaller than that of the less strongly bound Te2 or the Bi/Sb, thus it can be considered to act as the backbone for the structure[32,33]. The ADP ellipsoids are described by their MSD (mean square displacements, expressed in Å$^2$), the square root of which are the r.m.s. (root mean square displacements, expressed in Å), corresponding to their semi-axes: the major axis, labelled MSD$_{110}$, lies in-plane, making 30 degrees with the *a* and *b* directions (Fig. 2a), the medium axis (MSD$_{001}$) for all compositions and temperatures points out-of-plane, along *c*, and the minor in-plane axis, labelled MSD$_{110}$, then makes 60º with both *a* and *b*. The temperature dependence of the MSDs is shown in Fig. 2b for the optimized Bi$_{0.35}$Sb$_{1.65}$Te$_3$ compound, along with the analysis in terms of Einstein oscillators following the analysis of Ref. [29] using Eq. (1):

$$MSD_i = \frac{\hbar^2}{2m\theta_{E,i}} \coth \frac{\theta_{E,i}}{2k_BT} \qquad (1).$$

The characteristic energies, grouped around 4.6, 7, and 13.6 meV, agree well with those reported by inelastic neutron spectroscopy[34]. The analysis based on the MSDs highlights the specific atomic motions that correspond to each phonon mode. Although the diffraction based technique is inherently less accurate, based on a static measurement, it has the advantage of identifying the nature of the atoms (type and direction of motion) participating in each phonon mode. The lowest energies are 4.4 and 4.6 meV for in-plane vibration of Bi and Te2 (longest ellipsoid axis) and also 5.2 meV for the out-of-plane direction of Bi. These correspond to most favored vibrations given the distribution of chemical bonds across the structure: terminal Te2 is less weakly bonded given its interlayer position. It is believed that the rattling of these heavy atoms plays a paramount role on the relatively low thermal conductivity of these intermetallic materials. This rattling may be considered as a trap for phonons. The next group is 7.4 and 6.9 meV for Te1 and Te2, for the medium out-of-plane ellipsoid axis and also 6.7 meV for the Te1 the long axis. Finally, the largest energies are found for shorter in-plane motion (shortest axes) 13.3 and 13.8 meV for Bi and Te2 and 20.4 meV for Te1. The presence of the lone electron pairs lodged in the interlayer space accounts for the direction of the main vibration parallel to the layers, since the volume occupied by the electron pairs prevents or minimizes the vibrations out of the (001) planes, mainly for Bi and Te2 atoms. For Te1, showing the smallest ADPs, the formation of strong covalent bonds to the upper and lower layer of Bi atoms is responsible for the relatively contained vibrations, contributing in a minor way to the reduction of the thermal conductivity.

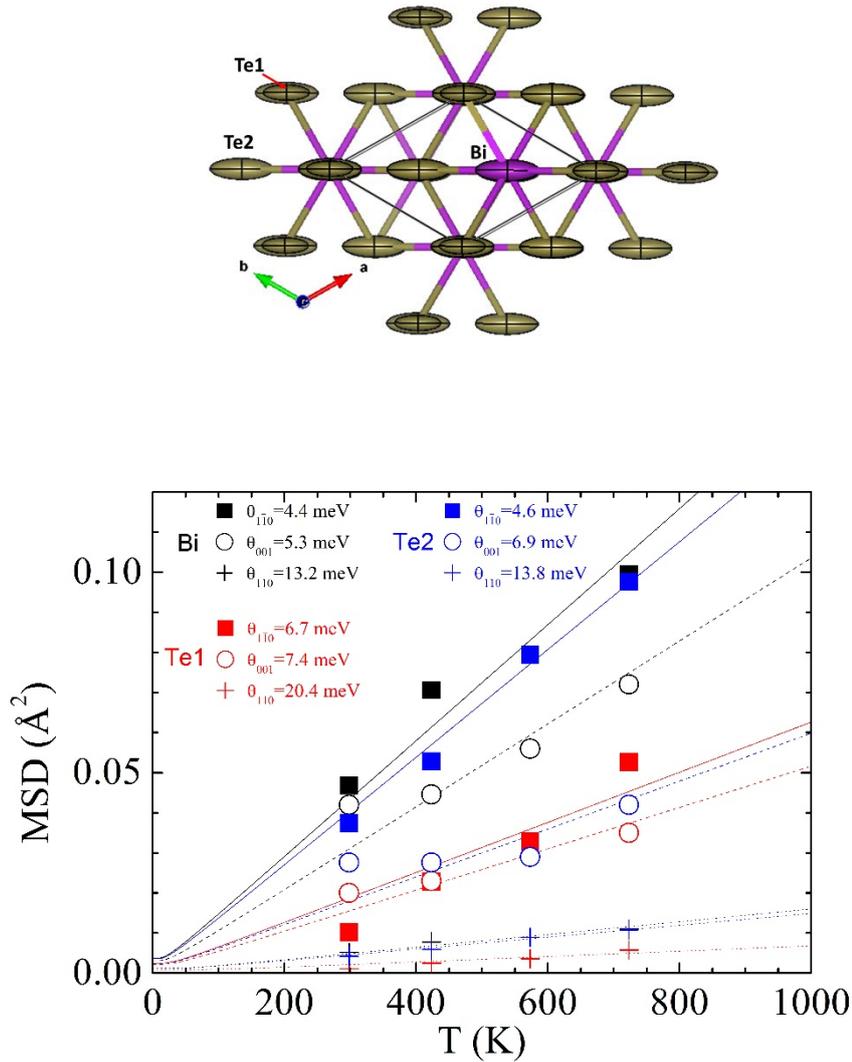

*Fig. 2 (Upper panel) Anisotropic displacement ellipsoids of $Bi_{0.35}Sb_{1.65}Te_3$ and (lower panel) their temperature dependence fitted to Einstein oscillators. The MSDs for Bi/Sb (black), Te1 (red) and Te2 (blue), represent the main axes, in-plane $MSD_{1\bar{1}0}$ (full squares and solid lines), $MSD_{110}$ (crosses and dotted lines), and out-of-plane $MSD_{001}$ (open*

*circles and dashed lines), respectively, of the displacement ellipsoids, and the resulting best fits. The labels indicate the best-fit Einstein oscillator energies according to Eq. (1).*

The increasing Sb content (x) has important effects on the crystallographic and vibrational properties that are not obvious from the static crystallographic parameters. The in-plane lattice parameter (*a*) and unit cell volume (V), in Fig. 3a, b decrease monotonously with the incorporation of more Sb, although the out-of-plane lattice parameter (*c*) hints at some non-monotonous changes near x=1.5. This is the result of a compromise between the smaller size of Sb *vs* Bi, tending to shrink the *c* axis as observed beyond x= 1.5 (Fig. 3c), and the increase of the covalent character of the Sb-Te bonds with respect to Bi-Te bonds. The higher covalence within the layers tends to decrease the Van der Waals interactions between adjacent layers, occurring mainly via terminal Te2 atoms; this effect predominates up to x=1 (Fig. 3c). The MSDs (Figs. 3c, d, e) also show an interesting phenomenology that deserves some comments. The longest axes of the Te2 ellipsoids defined by the MSDs corresponds to x= 0, but its length dramatically decreases upon Sb incorporation (Fig. 3d) and it approximately equals the long axis of the motion of (Bi,Sb) atoms. This trend is very interesting and again shows the increment of the covalency of (Bi,Sb)-Te bonds as x increases; the rattling effect of terminal Te2 atoms is reduced and at the same time (Bi,Sb) long axes grow due to the disordering, since both atoms are distributed at random at the same crystallographic sites. A similar effect is observed in the shorter axes (Fig. 3f), whereas the vibrations along the c axis (Fig. 3e) of the disc-shaped ellipsoids experience a non-monotonic evolution that may be related to the variation of the c unit-cell parameter attending to the same reasons above commented, acquiring a higher degree of freedom for large Sb contents.

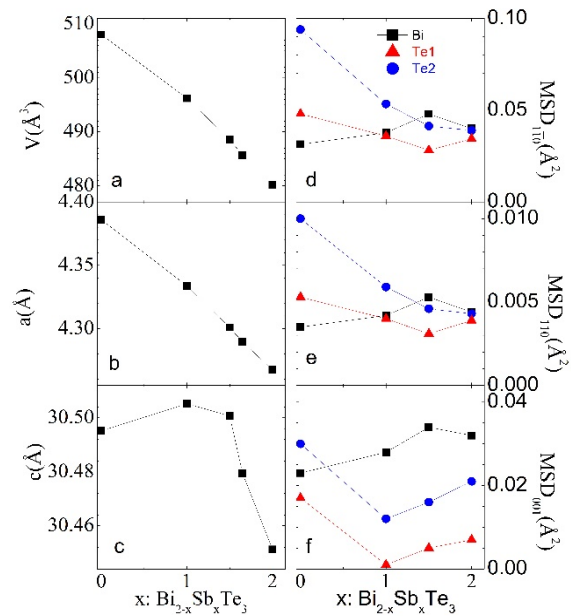

*Fig. 3 Composition dependence of the unit-cell volume (a) and lattice parameters (b, c) and of the MSDs ([1̄10], [110, [001] in d, e, f), respectively, at room temperature. Blue squares for Bi, green triangles for Te1 and red circles for Te2. MSD data for x= 1.65 are not included as they were collected in a different diffractometer.*

*Microstructure*

Arc-melting enhances the texture of the materials, producing stacked sheets, as shown by scanning electron microscopy (SEM) images in Fig. 4 for $Bi_{0.5}Sb_{1.5}Te_3$. The large surfaces are perpendicular to the *c*-axis, accounting for its easy cleaving. The thickness of the individual sheets is less than 50 nm. The thermoelectric properties of these materials are strongly influenced by this nanostructuration, providing many surface boundaries that bring about strong phonon scattering.

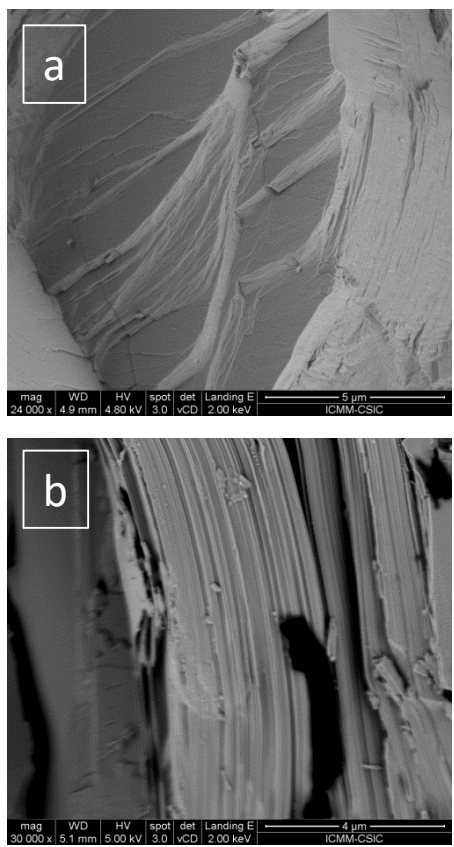

*Fig. 4 Scanning electron microscopy (SEM) images of as-grown $Bi_{0.5}Sb_{1.5}Te_3$ at (a) x2400 and (b) x30000 magnification. The samples are made up of nanometric platelets (less than 50 nm thick) normal to the c-axis.*

Nominal $Bi_{0.35}Sb_{1.65}Te_3$ crystals have been examined via electron diffraction combined with high resolution TEM. Diffraction patterns measured over long lateral distances exhibited ring-like features such as those of a polycrystal. An inhomogeneous distribution of crystal grain sizes was detected, with lateral dimensions mostly within the 1000-200 nm range. Still, the texture within these grains is not completely homogeneous. Figure 5a shows a low magnification image exhibiting an area approximately 90 x 60 nm in size, along with the corresponding diffraction pattern in Fig. 5b. Within such length scales (<100 nm), diffraction patterns exhibit features corresponding to a single crystal, in this case with a preferential orientation along the [001] zone axis, in this case. Other phases are also observed, but their volume fraction is low. A model of crystal structure based on the neutron data is shown in Fig. 5c. High resolution TEM images such as the one in Fig. 5d, which corresponds to the area highlighted with a yellow rectangle in Fig. 5a, exhibit a high degree of local crystallinity. Slight local misorientations within regions of tens of nm are detected (see the Fourier transforms corresponding to the areas highlighted in Fig. 5d). The nano-pocket marked with the red circle exhibits a [001] orientation. However, adjacent nano-regions (e.g. those marked with green or blue circles) display slight misalignments, as shown by the corresponding Fourier transforms on the right. These findings are consistent with the idea that the material exhibits an inhomogeneous crystal texture over lengths scales of tens of nm within the nanosheets seen in Fig. 4, which is favorable for thermoelectricity.

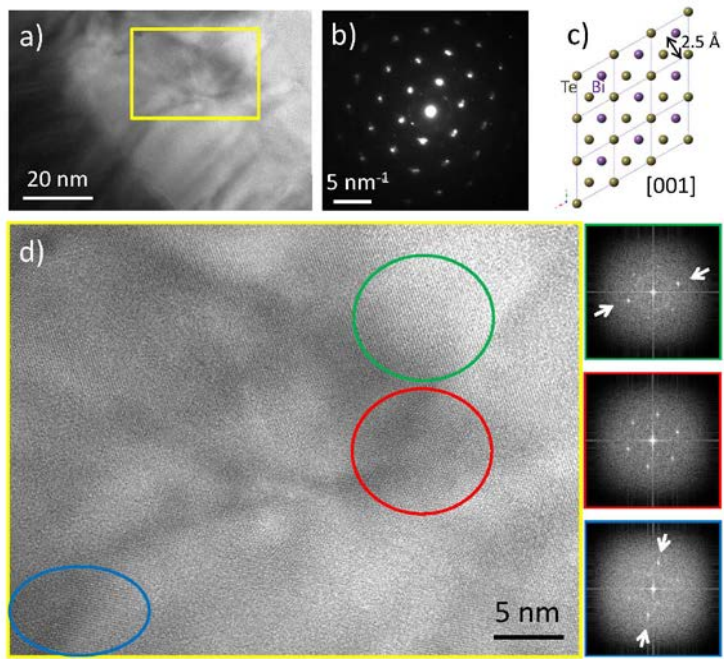

*Fig. 5 TEM study of a nominal $Bi_{0.35}Sb_{1.65}Te_3$ crystal. a) Low magnification bright field image b) along with the diffraction pattern measured over this area. Most of the grain is oriented down the [001] zone axis. c) Corresponding crystal model. d) High resolution zoom into the area highlighted with a yellow rectangle in a), where a number of varying crystal orientations can be detected. A red circle depicts an area with a [001] orientation, while the green and blue circles mark adjacent regions that are slightly off in terms of local orientation. Accordingly, the areas marked in green, red and blue exhibit different features in the Fourier transforms (panels on the right, from top to bottom respectively).*

*Electronic and thermal transport properties*

Our compounds present metallic behavior, showing the expected decrease of the electrical conductivity upon heating, consistent with their semimetallic nature (Fig. 6a) and in agreement with increased scattering on lattice vibrations. At 300 K, samples with composition $Bi_{0.5}Sb_{1.5}Te_3$ and $Bi_{0.35}Sb_{1.65}Te_3$ present resistivities of 28 and 35 µΩ.m, respectively, while for Bi-Sb-Te samples prepared by mechanical alloying and SPS, values in the 18- 30 µΩ.m range have been reported.[18,22,35] For chemically synthetized samples the resistivities are closer to 20 µΩ.m.[35] In spite of the strongly nanostructured texture of the pellets, the induced increase of the charge carrier scattering on grain boundaries and its effect on the electrical resistivity is relatively low.

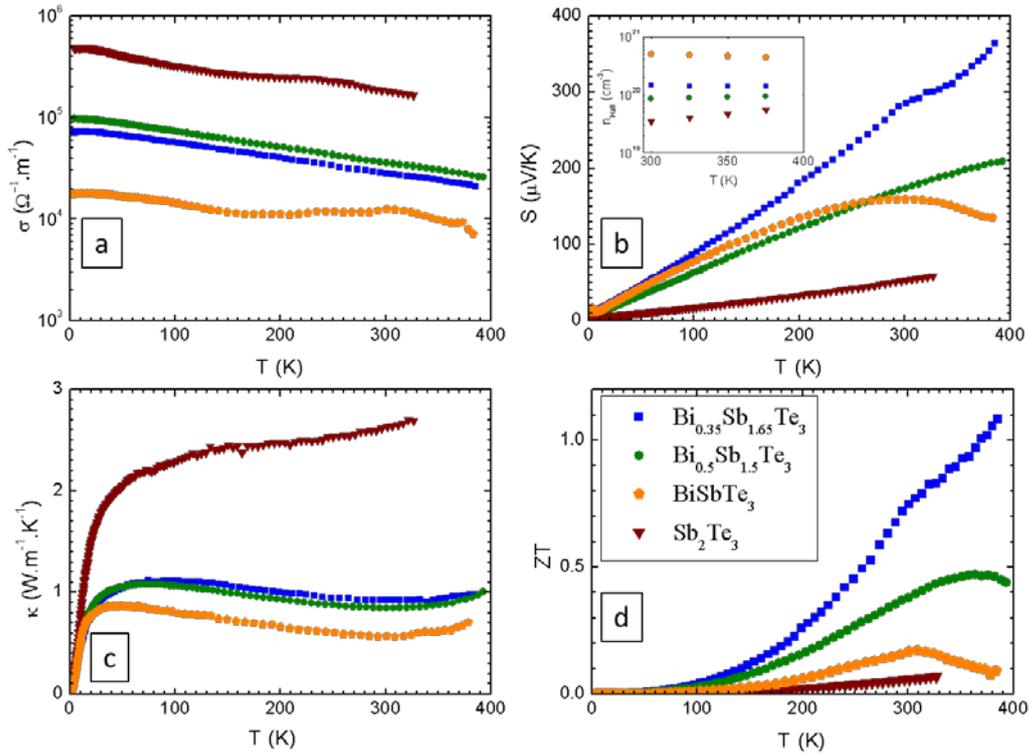

Fig. 6. Temperature dependent electronic and thermal transport properties of $Bi_{2-x}Sb_xTe_3$: (a) electrical conductivity, (b) Seebeck coefficient, (inset) charge carrier density from Hall-effect, (c) thermal conductivity, and (d) ZT figure of merit reaching 1.1 for the nominal composition $Bi_{0.35}Sb_{1.65}Te_3$.

The p-type Seebeck coefficients increase monotonously between 5 K and 300 K for all compositions (Fig. 6b), reaching a maximum of 209 µV K$^{-1}$ in $Bi_{0.5}Sb_{1.5}Te_3$ at 395 K, as checked in several samples. Pristine $Bi_2Te_3$ is an n-type semiconductor, with Seebeck-coefficients in the range -50 to -260 µV K$^{-1}$.$^{17,36,37}$ The literature indicates that as Bi$^{3+}$ is progressively replaced by Sb$^{3+}$, the sign of the conduction carriers becomes positive, reaching a maximum value 250 µV K$^{-1}$ at 320 K, $^{35,38}$ for bulk samples prepared by mechanical alloying and SPS, corresponding to a nominal composition of Bi$_{0.5}$Sb$_{1.5}$Te$_3$. Similar S values are found in chemically synthesized (sintered in Ar atmosphere) samples$^{39}$. In Bi$_{0.5}$Sb$_{1.5}$Te$_3$ prepared by ball-milling and hot-pressing$^{22}$ the Seebeck-coefficients at 300 K are close to 220 µV K$^{-1}$. Unfavourably for thermoelectricity, the increment of the Seebeck-coefficient with Sb doping also involves a decrease of the electronic conductivity, which is harmful for the thermoelectric performance of these materials. We aimed to optimize the Sb composition, looking for a compromise between S and σ. An additional sample prepared under the same conditions, with nominal composition Bi$_{0.35}$Sb$_{1.65}$Te$_3$ (determined composition *Bi$_{0.51(1)}$Sb$_{1.49(1)}$Te$_3$)*, showed a maximum value of S= 350 µV K$^{-1}$ at 395 K (Fig. 6b). This remarkably high Seebeck thermopower in the Bi-Sb-Te system is related to the strong nanostructuration and the low-energy electron filtering produced at grain boundaries. Also, in Bi$_2$Te$_3$ alloys an important source of charge carriers of different signs are antisite defects, as they are strongly affected by the composition and their concentration is hypothetically increased at the grain boundaries of the nanostructured materials.$^{22}$

The charge carrier density was determined between 300 K and 375 K from Hall-effect measurements using the relation n=-1/R$_H$e-, where R$_H$ is the Hall coefficient. We used 1 mm thick cold-pressed disks, in the resistivity option of the PPMS system up to 14 T magnetic field, using spring-loaded pins for contacts in an approximate van der Pauw geometry. The charge carrier concentration was found to be between 10$^{19}$—10$^{21}$ cm$^{-3}$ for the studied compositions (Fig. 6b inset). These values are comparable with those found in literature for the Bi-Sb-Te system.$^{24,25,27,35,39,40}$ We have calculated the electronic contribution to the thermal conductivity, using a Lorentz number of 1.6x10$^{-8}$ $^{27}$, and found that the lattice contribution at room temperature is about 85.3% in Bi$_{0.35}$Sb$_{1.65}$Te$_3$.

Arc-melting produces samples with low-thermal conductivity$^{28,41}$. We find the lowest κ= 0.56 W.m$^{-1}$.K$^{-1}$ at 309 K for BiSbTe$_3$, while at the same ambient temperature we find 0.84, 0.92 and 2.63 W.m$^{-1}$.K$^{-1}$ for Bi$_{0.5}$Sb$_{1.5}$Te$_3$, Bi$_{0.35}$Sb$_{1.65}$Te$_3$ and Sb$_2$Te$_3$ (Fig. 6c), respectively. These are among the best (lowest) values for the $Bi_2Te_3$ system,[22] with typical values of around 1.3 W.m$^{-1}$.K$^{-1}$ for commercial alloys. We postulate that it is the highly anisotropic, nanostructured, granular morphology that explains the useful discrepancy between high electrical and low thermal conductivities in these samples. The frequent grain (platelet) boundaries along the phonon paths[28] lead to strong phonon scattering, as the platelet thickness is probably comparable to the phonon diffusion length. Yet, the electrical conductivity is much less affected because of the large aspect ratio of the platelets and the different length scale for electron scattering. The arc-melting involves a very rapid quenching of the molten material in the water cooled copper crucible. Nanostructuring has been demonstrated to lead to lower thermal conductivity in $Bi_{0.5}Sb_{1.5}Te_3$ obtained by different physicochemical procedures, via ball-milling and hot-pressing, of 1.0 W.m$^{-1}$.K$^{-1}$ at 370 K, or in SPS sintered samples, between 0.7 and 1.2 W.m$^{-1}$.K$^{-1}$ at 325 K. [18,22,35,40] Eventually, for the optimized composition of $Bi_{0.35}Sb_{1.65}Te_3$ (with actual composition *$Bi_{0.51}Sb_{1.49}Te_3$*), the thermoelectric figure of merit increases above ZT=1.1 (Fig. 6d) at the limiting temperature (395 K) of the PPMS instrument.

**Conclusions**

In summary, the present structural study from neutron data contributes with important insights to the transport properties relevant for thermoelectricity in bismuth-telluride. The thermal conductivity is to a large extent governed by the characteristic phonon energies associated to each type of atom in the crystallographic unit, which were determined from the anisotropic displacement ellipsoids across the series: subtly affected by the increase of covalency upon Sb introduction. Furthermore, we obtained a thermoelectric figure of merit of ZT>1.1 for an optimized nominal composition ($Bi_{0.35}Sb_{1.65}Te_3$) with an exceptional Seebeck coefficient of S= 350 µV K$^{-1}$ at 395 K. Although higher ZT values have been reported in p-type $Bi_2Te_3$ systems prepared by complex elaboration protocols,[27] arc-melting has the virtues of being straightforward, fast and cost-effective, compared to other methods such as ball-milling which requires several hours and additional synthesis and compaction steps[22], whereas the energy consumption of the arc furnace working for several seconds is comparatively marginal. It produces polycrystalline, strongly nanostructured ingots. It leads to minimized thermal conductivity while preserving large electronic conductivity, through enhanced phonon scattering. Nanostructuration also accounts for excellent Seebeck coefficients in robust, easy to handle pellets that could be directly integrated into thermoelectric devices.

**Methods**

The details of the sample preparation, structural and transport property characterizations were given elsewhere[28,29,41]; we summarize them here. Stoichiometric mixtures of the elements comprising the alloys of $Bi_{2-x}Sb_xTe_3$ (x= 0.0, 1.0, 1.5, 1.65, 2.0) were melted in an Edmund Buhler MAM-1 furnace under Ar atmosphere. The resulting pellets were then prepared for structural or physical characterization. For structural studies, powders were prepared by grinding in a mortar. Neutron powder diffraction (NPD) data were collected at room temperature for x= 0.0, 1.0, 1.5, and 2.0 at the HRPT diffractometer of the SINQ spallation source (Switzerland), with a wavelength λ= 1.494 Å, during 2 h of acquisition time in the high-intensity mode, using rotating vanadium cylinder sample holders (dia. 8 mm). For the optimized composition with x= 1.65, temperature-dependent NPD data were collected at the D2B diffractometer of the Institut Laue-Langevin, with a neutron wavelength of λ= 1.594 Å in the high-flux configuration. Approximately 2 g of the sample were enclosed in a vanadium holder and placed in a furnace functioning under vacuum (P ≈ $10^{-6}$ torr). The measurements were carried out at 423, 573 and 723 K during the warming run. In all cases, the crystal structures were refined by the Rietveld method using the FULLPROF refinement program.[42] We used the following coherent scattering lengths for Bi, Sb and Te, respectively: 8.532, 5.570 and 5.80, fm. The calculation was corrected assuming a preferred orientation defined in the [001] direction, considering the formation of platelets perpendicular to the Cu crucible of the arc furnace. We measured simultaneously the Seebeck coefficient, electrical and thermal conductivities, in a Physical Properties Measurements System (PPMS) in a vacuum of $10^{-5}$ Torr with residual He atmosphere, in the temperature range of 2 to 380 K. For this, we pressed, under uniaxial pressure of 70 bars, the as-grown pellets into 10x3x2 $mm^3$ bars with perfectly parallel faces. The measured density was always higher than 95% of the crystallographic density. We attached four Cu wires with silver epoxy. We applied a constant temperature gradient of 3% across the sample during Seebeck and thermal conductivity measurements. High resolution transmission electron microscopy (TEM) observations have been carried out in a JEOL 2100 TEM, operated at 200 kV. Polycrystalline powder has been diluted in ethanol and dispersed in the ultrasonic device. The cationic ratios were determined by inductively coupled plasma-atomic emission spectroscopy (ICP-AES) ICP PERKIN ELMER mod. OPTIMA 2100 DV equipment. The samples were dissolved in hydrochloric plus nitric acid and then diluted with distilled water.

**References**

1. Martín-González, M., Caballero-Calero, O. & Díaz-Chao, P. Nanoengineering thermoelectrics for 21st century: Energy harvesting and other trends in the field. *Renew. Sustain. Energy Rev.* **24,** 288–305 (2013).

2. Snyder, G. J. & Toberer, E. S. Complex thermoelectric materials. *Nat. Mater.* **7,** 105–114 (2008).

3. Heikes, R. R. & Ure, R. W. *Thermoelectricity: Science and Engineering*. (Interscience, 1961).

4. Rosi, F. D. Thermoelectricity and thermoelectric power generation. *Solid. State. Electron.* **11,** 833–848 (1968).

5. Rosi, F. D., Hockings, E. F. & Lindenblad, N. E. Semiconducting materials for thermoelectric power generation. *RCA Rev.* **22,** 82–121 (1961).

6. Wood, C. Materials for thermoelectric energy conversion. *Rep. Prog. Phys* **459,** 459–539 (1988).

7. Goldsmid, H. J. & Douglas, R. W. The use of semiconductors in thermoelectric refrigeration. *British Journal of Applied Physics* **5,** 458–458 (2002).

8. Nolas, G. S., Sharp, J. & Goldsmid, H. J. *Thermoelectrics: Basic Principles and New Materials Developments*. *Book* (2001).

9. Scherrer, H. & Scherrer, S. *Thermoelectrics Handbook Macro to Nano*. (CRC, 2006).

10. G. Chen, M. S. Dresselhaus, G. Dresselhaus, J.-P. Fleurial, T. C. Recent developments in thermoelectric materials. *Int. Mater. Rev.* **48,** 45 (2003).

11. Dresselhaus, M. S. *et al.* New directions for low-dimensional thermoelectric materials. *Adv. Mater.* **19,** 1043–1053 (2007).

12. Lin, Y. M., Rabin, O., Cronin, S. B., Ying, J. Y. & Dresselhaus, M. S. Semimetal-semiconductor transition in Bi1-xSbx alloy nanowires and their thermoelectric properties. *Appl. Phys. Lett.* **81,** 2403–2405 (2002).

13. Rabin, O., Lin, Y. M. & Dresselhaus, M. S. Anomalously high thermoelectric figure of merit in Bi1-x-Sbx nanowires by carrier pocket alignment. *Appl. Phys. Lett.* **79,** 81–83 (2001).

14. Amy L. Prieto, Melissa S. Sander, Marisol S. Martın-Gonzalez, Ronald Gronsky, Timothy Sands, A. M. S. Electrodeposition of Ordered Bi2Te3 Nanowire Arrays. *J. Am. Chem. Soc.* **123,** 7160–7161 (2001).

15. Heremans, J. *et al.* Bismuth nanowire arrays: Synthesis and galvanomagnetic properties. *Phys. Rev. B* **61,** 2921–2930 (2000).

16. Venkatasubramanian, R., Siivola, E., Colpitts, T. & O'Quinn, B. Thin-film thermoelectric devices with high room-temperature figures of merit. *Nature* **413,** 597–602 (2001).

17. Lan, Y., Minnich, A. J., Chen, G. & Ren, Z. Enhancement of Thermoelectric Figure-of-Merit by a Bulk Nanostructuring Approach. *Adv. Funct. Mater.* **20,** 357–376 (2010).

18. Bomshtein, N., Spiridonov, G., Dashevsky, Z. & Gelbstien, Y. Thermoelectric , Structural , and Mechanical Properties of Spark-Plasma-Sintered Submicro- and Microstructured p-Type Bi0.5Sb1.5Te3. *J. Electron. Mater.* **41,** 1546–1553 (2012).

19. Cao, Y. Q., Zhao, X. B., Zhu, T. J., Zhang, X. B. & Tu, J. P. Syntheses and thermoelectric properties of Bi2 Te3 Sb2 Te3 bulk nanocomposites with laminated nanostructure. *Appl. Phys. Lett.* **92,** 143106 (2008).

20. Zhao, Y., Dyck, J. S., Hernandez, B. M. & Burda, C. Enhancing thermoelectric performance of ternary nanocrystals through adjusting carrier concentration. *J. Am. Chem. Soc.* **132,** 4982–4983 (2010).

21. Pettes, M. T. *et al.* Thermoelectric transport in surface- and antimony-doped bismuth telluride nanoplates Thermoelectric transport in surface- and antimony-doped bismuth telluride nanoplates. **104810,** 1–10 (2016).

22. Poudel, B. *et al.* High-Thermoelectric Performance of Nanostructured Bismuth

Antimony Telluride Bulk Alloys. *Science* **320,** 634–638 (2008).

23. Bochentyn, B., Miruszewski, T., Karczewski, J. & Kusz, B. Thermoelectric properties of bismuth e antimony e telluride alloys obtained by reduction of oxide reagents. *Mater. Chem. Phys.* **177,** 353–359 (2016).

24. Zhang, C., Peng, Z., Li, Z. & Yu, L. Controlled growth of bismuth antimony telluride Bi x Sb 2 À x Te 3 nanoplatelets and their bulk thermoelectric nanocomposites. *Nano Energy* **15,** 688–696 (2015).

25. Luo, Y. *et al.* Melting and solidi fi cation of bismuth antimony telluride under a high magnetic fi eld : A new route to high thermoelectric performance. *Nano Energy* **15,** 709–718 (2015).

26. Hu, L. *et al.* Shifting up the optimum figure of merit of p -type bismuth telluride-based thermoelectric materials for power generation by suppressing intrinsic conduction. **6,** e88-8 (2014).

27. Kim, S. Il *et al.* Dense dislocation arrays embedded in grain boundaries for high-performance bulk thermoelectrics. *Science* **348,** 109–114 (2015).

28. Serrano-Sánchez, F. *et al.* Record Seebeck coefficient and extremely low thermal conductivity in nanostructured SnSe. *Appl. Phys. Lett.* **106,** 83902 (2015).

29. Gharsallah, M. *et al.* Nanostructured Bi 2 Te 3 Prepared by a Straightforward Arc-Melting Method. *Nanoscale Res. Lett.* **11,** 142 (2016).

30. Adam, A. Rietveld refinement of the semiconducting system Bi2−xFexTe3 from X-ray powder diffraction. *Mater. Res. Bull.* **42,** 1986–1994 (2007).

31. Shannon, R. D. Revised Effective Ionic Radii and Systematic Studies of Interatomie Distances in Halides and Chaleogenides. *Acta Cryst.* **32,** 751 (1976).

32. Falmbigl, M. *et al.* Type-I clathrate Ba8Ni(x)Si(46-x): phase relations, crystal chemistry and thermoelectric properties. *Dalton Trans.* **41,** 8839–49 (2012).

33. Paschinger, W. *et al.* Ba-filled Ni-Sb-Sn based skutterudites with anomalously high lattice thermal conductivity. *Dalton Trans.* **45,** 11071–11100 (2016).

34. Rauhi, H., Geickt, R., Kohleri, H., Nuckers, N. & Lehnerd, N. Generalised

phonon density of states of the layer compounds Bi2Se3, Bi2Te3, Sb2Te3 and Bi2(Te0.5Se0.5)3, (Bi0.5Sb0.5)2Te3. *J. Phys. C Solid State Phys.* **14,** 2705–2712 (1981).

35. Chen, C., Liu, D. W., Zhang, B. P. & Li, J. F. Enhanced thermoelectric properties obtained by compositional optimization in p-type Bi x Sb 2-x Te 3 fabricated by mechanical a35 lloying and spark plasma sintering. *J. Electron. Mater.* **40,** 942–947 (2011).

36. Saleemi, M., Toprak, M. S., Li, S., Johnsson, M. & Muhammed, M. Synthesis, processing, and thermoelectric properties of bulk nanostructured bismuth telluride (Bi 2 Te 3 ). *J. Mater. Chem.* **22,** 725–730 (2012).

37. Peranio, N. *et al.* From thermoelectric bulk to nanomaterials: Current progress for Bi 2 Te 3 and CoSb 3. *Phys. Status Solidi A* **11,** 1–11 (2015).

38. Kitagawa, H., Nagao, K., Mimura, N., Morito, S. & Kikuchi, K. Preparation of Bi0.5Sb1.5Te3 thermoelectric materials by pulse-current sintering under cyclic uniaxial pressure. *J. Electron. Mater.* **43,** 1574–1579 (2014).

39. Dyck, J. S., Mao, B., Wang, J., Dorroh, S. & Burda, C. Effect of Sintering on the Thermoelectric Transport Properties of Bulk Nanostructured Bi0.5Sb1.5Te3 Pellets Prepared by Chemical Synthesis. *J. Electron. Mater.* **41,** 1408–1413 (2012).

40. L. Han, S.H. Spangsdorf, N.V. Nong, L.T. Hung, H.P. Ngan, Y. Chen, A. Roch, L. Stepien, and N. P. Effects of Spark Plasma Sintering Conditions on the Anisotropic Thermoelectric Properties of Bismuth Antimony Telluride. *RSC Adv.* **6,** 59565–59573 (2016).

41. Gharsallah, M. *et al.* Giant Seebeck effect in Ge-doped SnSe. *Sci. Rep.* **6,** 26774 (2016).

42. Rodríguez-Carvajal, J. Recent advances in magnetic structure determination by neutron powder diffraction. *Phys. B* **192,** 55–69 (1993).

**Acknowledgement**

We are grateful to the Spanish Ministry of Economy and Competitivity for granting the project MAT2013-41099-R, and to PSI and ILL for making all facilities available for the neutron diffraction experiments. Electron microscopy observations carried out at the UCM Centro Nacional de Microscopía electronica. Financial support from ERC PoC MAGTOOLS and Spanish MINECO/Feder MAT2015-66888-C3-3-R is acknowledged.

**Author Contributions**

FSS and MG synthesized the samples. FSS and NNM carried out the transport experiments. FSS, MTFD and JAA performed the structural characterization work. NB and MV carried out the TEM study. JLM, NNM and JAA conceived of the study. All authors participated in discussing the results and commented on the manuscript.

**Data availability statement**

Raw experimental data and samples are available upon request to the authors.

**Additional Information**

**Competing financial Interests**: The authors declare no competing financial interests.

Table 1. Unit-cell parameters and main interatomic distances in the $Bi_{2-x}Sb_xTe_3$ system from NPD data.

|  | $Bi_2Te_3$* | Bi SbTe$_3$ | $Bi_{0.5}Sb_{1.5}Te_3$ | $Bi_{0.35}Sb_{1.65}Te_3$ | $Sb_2Te_3$ |
|---|---|---|---|---|---|
| **a (Å)** | 4.3859(2) | 4.3337(1) | 4.3008(1) | 4.2894(6) | 4.2673(2) |
| **c (Å)** | 30.495(2) | 3.5052(2) | 30.5007(2) | 30.4795(1) | 30.451(2) |
| **V (Å³)** | 508.03(5) | 496.21(4) | 488.59(4) | 485.67(3) | 480.21(4) |
| **Te1-Bi/Sb x3** | 3.253(4) | 3.227(3) | 3.204(4) | 3.185(1) | 3.162(6) |
| **Te2-Bi/Sb x6** | 3.061(5) | 3.024(4) | 3.000(4) | 2.999(1) | 2.995(6) |
| **Te2-Te2 x3** | 3.660(6) | 3.678(4) | 3.706(4) | 3.717(3) | 3.715(5) |
| **Te2-Te2 x6** *inter-layer* | 4.3859(1) | 4.3337(1) | 4.3008(1) | 4.289(6) | 4.2673(1) |

*From ref [29]